\documentclass[twocolumn,prl]{revtex4}

\begin{document}


\title{Sandpile formation by revolving rivers}

\author{
E. Altshuler$^1$,
O. Ramos$^{1,2}$,
A.J. Batista-Leyva$^{1,3}$,
A. Rivera$^{4}$
and K.E.  Bassler$^{5}$}

\affiliation{ ~$^1$Superconductivity Laboratory, IMRE-Physics
Faculty, University of Havana, 10400 Havana, Cuba \\ ~$^2$Physics
Department, ELAM, 19100 Havana, Cuba\\ ~$^3$Physics Department,
University of Holgu\'{\i}n, Holgu\'{\i}n 80100, Cuba\\ ~$^4$
Zeolite Engineering Laboratory, IMRE-Physics Faculty, University
of Havana, 10400 Havana, Cuba\\ ~$^5$Department of Physics,
University of Houston, Houston, TX 77204-5005}

\date{\today}

\begin{abstract}
Experimental observation of a new mechanism of sandpile formation
is reported. As a steady stream of dry sand is poured onto a horizontal
surface, a pile forms which has a thin river of sand on one side
flowing
from the apex of the pile to the edge of its base.
The river rotates about the pile, depositing
a new layer of sand with each revolution, thereby growing the
pile. For small piles the river is
steady and the pile formed is smooth. For larger piles, the river
becomes intermittent and the surface of the pile becomes undulating.
The frequency of revolution of the river is measured as the pile grows
and the results are explained with a simple scaling argument.
The essential features of the system that produce the phenomena are
discussed.
\end{abstract}


\maketitle


Sandpiles have received considerable interest because of their
intrinsic scientific interest both from the fundamental and
applied points of view, and also because they are simple examples
of complex systems whose behavior has been used in an attempt to
explain a variety of physical, chemical, biological and social
phenomena\cite{B96}. Conventional understanding of sandpile
formation is that as grains of sand are poured onto a horizontal
surface, a conical pile develops which grows intermittently
through avalanches that ``adjust'' the angle of repose of the pile
about some critical value, or, at least, keep it between two
critical values. This mechanism of pile formation has been widely
studied in the recent years\cite{BTW87,JLN88,HSK90,RVK93,RVK94,M94,BCRE95,F96,CCF96,BRG98,DD99,DM00,TCB00,ARM01,AT01,ARG02}.
Here we report experimental observation of a remarkable new
mechanism of pile formation.

Pouring a steady stream of sand into the center of a cylindrical
container, as shown in Fig.~1, a pile formed. Then, a continuous
river of sand developed flowing from the apex of the pile to the
inner boundary of the container. The river, which was narrow
compared with the radius of the container, revolved around the
pile depositing a helical layer of sand a few grains thick with
each revolution. Thus, the pile grew as the river revolved around
it. A photograph of a revolving river can be seen in Fig.~2.
Within a range of experimental parameters and conditions, the
formation of a revolving river was easily reproducible, and very
robust. Once formed, a typical river persisted for dozens of full
turns around the growing pile, and stopped only when forced to by
interrupting the pouring of sand.

In the experiments, a vertical glass tube with a 20 mm inner
diameter was initially filled with sand using a funnel. Then a 4
mm hole was opened in the bottom of the tube, allowing sand to
fall out of the tube by its own weight. This arrangement produced
a steady flow of sand out of the tube at a steady rate of 4.5 g/s
for the duration of the experiment. Video cameras recorded both
lateral and top views of the piles during the experiment. (Top
views were obtained with the help of a $45^o$ tilted mirror). Two
different versions of the experiment were performed, each
corresponding to a different boundary condition of the growing
pile. In the first version (described above), the pile had a {\it
closed boundary}. The sand was dropped at the center of a
cylindrical container, so that the radius of the resulting pile
was constant in time. In the second version, the pile had an {\it
open boundary}. No container was present. Instead, the sand fell
onto a flat horizontal surface and the radius of the pile
increased in time.

Rivers that revolved about the pile in both clockwise and
counterclockwise directions were observed. The direction chosen in
a particular case depended on the initial conditions. The axial
symmetry of the system was therefore spontaneously broken as the
river was formed. Viewed from above, the rivers were slightly
bent, and always revolved around the pile in the direction of
their concavity, as shown in Figs. 2a and 2c. A steady revolving
river was typically observed when sand was poured into a container
with a 4-6 cm radius. In this case, the surface of the pile was
smooth. However, when a container with a radius larger than 6 cm
was used, an instability appeared in the flow of the river. The
revolving river still developed, growing the pile as before, but
the flow of the river was intermittent rather than continuous. In
that case, the intermittent flow produced an undulating pattern on
the pile surface, visible in Figs. 2c and 2d. The undulating
pattern resembles those recently observed for rapid granular flows
on an inclined plane\cite{FP01,AT02}, but presumably is caused by
a different mechanism. The observed pattern was quite regular for
containers with a radius just large enough to observe the
instability, but became more irregular as the size of the
container grew. If a container smaller than 3 cm radius was used,
stable revolving rivers were not observed.

The revolving river mechanism of pile
formation has also been observed by simply pouring the sand onto a
flat surface. In that case, the crossover from a continuously
flowing revolving river, observed in smaller piles, to an
intermittently flowing river, observed in larger piles, occurred
as the radius of the pile reached about 6 cm. The crossover
appeared to be correspond to the pile size needed for the length
of the ballistic motion of the sand grains in a river to begin to
be damped.

We have varied the drop height in the experiment.  For drop heights
between 1 cm and 7 cm the results closely follow the description given
above.
However, for drop heights less than 1 cm or larger than 7 cm,
stable rivers were not observed.

The origin of the curved shape of a revolving river and the reason
for it moving in the direction of its concavity can be understood
by how a river forms. Based on careful observation, revolving
rivers appear to form through the following scenario, illustrated
in Fig.~3. Initially, sand is poured onto the top of a conical
pile and it forms a river flowing straight down one side of the
pile (Fig.~3a). Sand begins to build up at the bottom of the river
at the edge of the pile, forming a growing inverted V shaped delta
of stationary sand (Fig.~3b). The delta grows in size until the
river spontaneously chooses to begin to flow down one of the sides
of the delta (Fig.~3c). Once it chooses a side, it continues to
flow down that side of the delta, depositing sand all along the
lower, delta side of the river. As it does so, it rotates about
the pile. For smaller piles, the process of rotation was stable.
However, for larger piles, it was not. Instead, in that case, a
new delta would intermittently begin to form at the bottom of the
river (Fig.~3d). When the delta reached sufficient height, the
river would ``jump'' forward in its rotation, and then begin
forming yet another new delta.

In order to begin to quantitatively understand the revolving
rivers, we measured the time evolution of the angular velocity of
river rotation with both closed and open boundaries. As shown in
Fig.~4, the angular velocity of river rotation was roughly
constant for piles in cylindrical containers, while it decreased
in time as $t^{-\alpha}$, with $\alpha = 2/3$, for open boundary
conditions. These results can be explained using the following
scaling argument whose geometrical hypotheses are illustrated in
Fig. 5. Assume that a new layer of sand is uniformly deposited on
a conical pile of radius $r$ with an angle of repose $\theta_c$,
and that the volume of sand added per unit time is $F$. For a
system with a closed boundary, the thickness of an added layer is
proportional to $\delta h$ (see Fig.~5a). Therefore, the volume of
sand deposited in each rotation of the river
$$ V = {\pi \; r^2 \; \delta h \over \cos \theta_c } $$
is
constant in time. The angular velocity of the river
\begin{equation}
\omega = 2\pi {F \over V}
\end{equation}
is therefore also constant in time. In our experiments, we
measured $\delta h =$ 2 mm, $\phi_c = 33^o$, and $F=0.35$
cm$^{3}$/s . However, for a system with an open boundary the
radius of pile grows in time. In this case, the thickness of each
layer is proportional to $\delta r$ (see Fig.~5b). The volume of
sand deposited in a rotation of the river is
\begin{equation}
V = \pi \;
\tan \theta_c \;
r^2 \; \delta r
\end{equation}
where $r$ is a function of time, but $\delta r= \delta h/\sin \theta_c$ is
constant.
Thus, from this result and Eqn.~1, $\omega \sim r^{-2}$. The pile radius
increases
at a rate of
$$
{dr \over dt}
=
\omega \; \delta r
$$
Integrating this expression, we get $r \sim t^{1/3}$, and therefore
$$
\omega \sim t^{-2/3} \; .
$$
Our scaling argument matches well the experimental results shown
in Fig.~4a. In the case of Fig.~4b, although this argument
correctly predicts the scaling of the experimental data for larger
piles, it does not properly describe the behavior of smaller
piles, presumably due to the fact that our geometrical assumptions
are inaccurate near the tip of the pile.

The appearance of revolving rivers is quite sensitive to the type of
sand used in the experiments.
In the results reported here, sand from Santa Teresa, Cuba, was
used. It consists in irregularly shaped grains of size $30-250 \; {\mu} m$
made of almost pure silicon oxide. It was also quite dry. Revolving rivers
were still observed if the sand was meshed to remove grains smaller than
$90\;{\mu} m$ and larger than $160 \; {\mu}m$. However, other sands from Cuba,
USA,
Norway and Tunisia were tried, including ones high in Calcium Carbonate,
and ones high in Magnetite, but no revolving rivers were observed within
our experimental conditions (a river occasionally formed in those sands,
but it disappeared in fractions of a second). Revolving rivers also were
not
observed if glass beads having roughly the same size as the Santa Teresa
sand
were used. It is therefore suspected that the effective coefficient of
friction between grains, and the mass density of grains may be
important factors determining if revolving rivers appear in the formation
of piles. These elements must be included in a future ``first principles''
model of the revolving rivers.

We acknowledge E. Mart\'{\i}nez for cooperation in the experiments, and
S. Douady, H. Herrmann, T. H. Johansen, R. Mulet,O. Pouliquen, H. Seidler,
O. Sotolongo and J.E. Wesfried for useful discussions and comments. We
thank
material support from the University of Havana's "Alma Mater" grants
programme. KEB acknowledges support from
the NSF through grant \#DMR-0074613, from
the Alfred P. Sloan Foundation, and from the Texas Center for
Superconductivity.




\newpage
~
\newpage


\subsection*{Figure Captions}

\vspace{0.2truein}

Fig. 1. Experimental setup

\vspace{0.2truein}

Fig. 2. Formation of a pile of sand by revolving rivers. The sand
is poured vertically on the center of cylindrical containers with
flat, horizontal bottoms at a deposition rate of 0.35 $cm^3/s$,
from a constant height of 1.5 cm above the apex of the pile. (a)
Top view of a pile growing into a 5 cm radius container, where the
continuous river can be identified. (b) Lateral view of the pile
shown in (a). (c) Top view of a pile growing into a 10 cm radius
container where an intermittent river and the related pattern can
be identified. (d) Lateral view of the pile shown in (c) (the
photo shows about 3 cm of the container's perimeter). In all
cases, arrows indicate the revolving direction.

\vspace{0.2truein}

Fig. 3. Development of a revolving river. (a) A river flows straight down
the side of the pile, and a delta begins to form at its bottom. (b) The
delta continues to grow. (c) When the delta is sufficient size, the river
begins to flow down one side and rotate around the pile. (d) If the pile
is sufficiently large, a new delta forms intermittently at the bottom of
the
river, causing the rotation of the river to become intermittent.

\vspace{0.2truein}

Fig. 4. Time dependence of the angular speed of revolving rivers for (a)
closed boundary conditions in the continuous regime(5 cm-radius container)
and (b) open boundary conditions. The solid line in (b) has a slope
of -2/3.

\vspace{0.2truein}

Fig. 5. Geometrical hypotheses of our scaling argumet for (a)
closed boundary conditions and (b) open boundary conditions.


\begin{references}
\bibitem{B96} For example, see P. Bak,
 {\it How Nature Works: The Science of
Self-Organized Criticality} (Copernicus, New York, 1996).

\bibitem{BTW87} P. Bak, C. Tang and K. Wiesenfeld,
Phys. Rev. Lett. {\bf 59}, 381 (1987), Phys. Rev. A {\bf 38}, 364 (1988).

\bibitem{JLN88} H. M. Jaeger, Ch.-H. Liu, and S. R.
Nagel. Phys. Rev. Lett. {\bf 62}, 40 (1988).

\bibitem{HSK90} G.A. Held, D.H. Solina, D.T. Keane, W.J.Haag,
P.M. Horn, and G. Grinstein, Phys. Rev. Lett. {\bf 65}, 1120 (1990).

\bibitem{RVK93} J. Rosendahl, M. Vekic, and J.
Kelley, Phys. Rev. E {\bf 47}, 1401 (1993).

\bibitem{RVK94} J. Rosendahl, M. Vekic and J.E.
Rutledge, Phys. Rev. Lett. {\bf 73}, 537 (1994).

\bibitem{M94} A.Mehta (ed.) Granular Matter (Springer, Heidelberg,
1994).

\bibitem{BCRE95} J.-P. Bouchaud, M. E. Cates, J. Ravi Prakash, and S. F.
Edwards,
Phys. Rev. Lett. {\bf 74}, 1982 (1995).

\bibitem{F96} V. Frette, {\it et al}. Nature, {\bf 379},
49 (1996).

\bibitem{CCF96} K. Christensen, A. Corral, V. Frette,J.
Feder and T. J{\o}ssang, Phys. Rev. Lett. {\bf 77} 107 (1996).

\bibitem{BRG98} T. Boutreux, E. Rapha\"el, and P.-G. de Gennes,
Phys. Rev. E {\bf 58}, 4692 (1998).

\bibitem{DD99} A. Daerr and S. Douady, Nature {\bf 399}, 6733
(1999).

\bibitem{DM00} S. N. Dorogovtsev, and J. F. F. Mendes, Phys. Rev. E {\bf
61},
2909 (2000).

\bibitem{TCB00} E. Thorsten, P. Claudin, and J. P. Bouchaud, Europhys.
Lett.
{\bf 50},
594 (2000).

\bibitem{ARM01} E. Altshuler, O. Ramos, C. Mart\'{\i}nez, L.
E. Flores and C. Noda. Phys. Rev. Lett. {\bf 86}, 5490 (2001).

\bibitem{AT01} I. S. Aranson and L. S. Tsimring, Phys. Rev. E {\bf 64},
020301 (2001).

\bibitem{ARG02} A. Aradian, E. Raphael, P.-G. de Gennes,
C. R. Physique {\bf 3}, 187 (2002).

\bibitem{FP01} Y. Forterre and O. Pouliquen, Phys.
Rev. Lett. {\bf 86}, 5886 (2001).

\bibitem{AT02} I. S. Aranson and L. S. Tsimring, Preprint.
cond-mat/0203189
(2001).

\end{references}
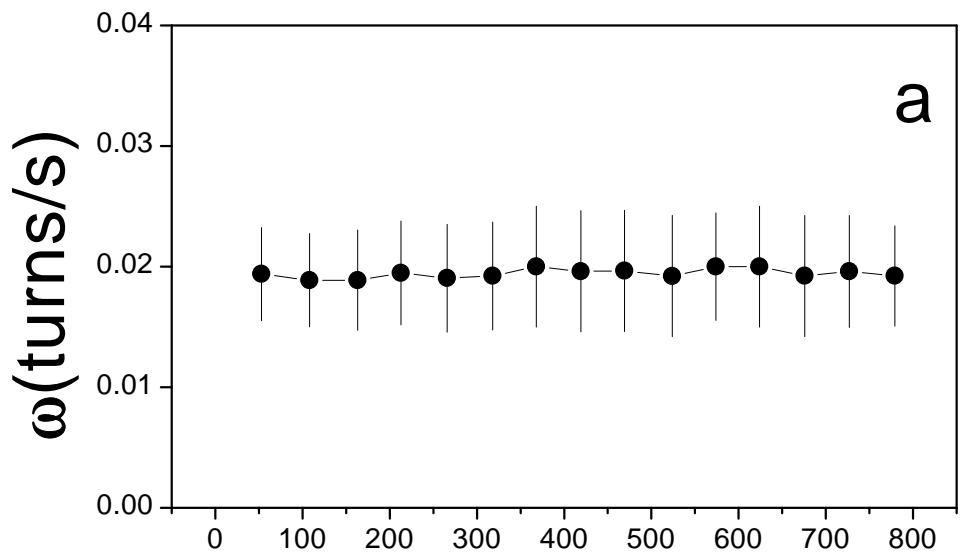
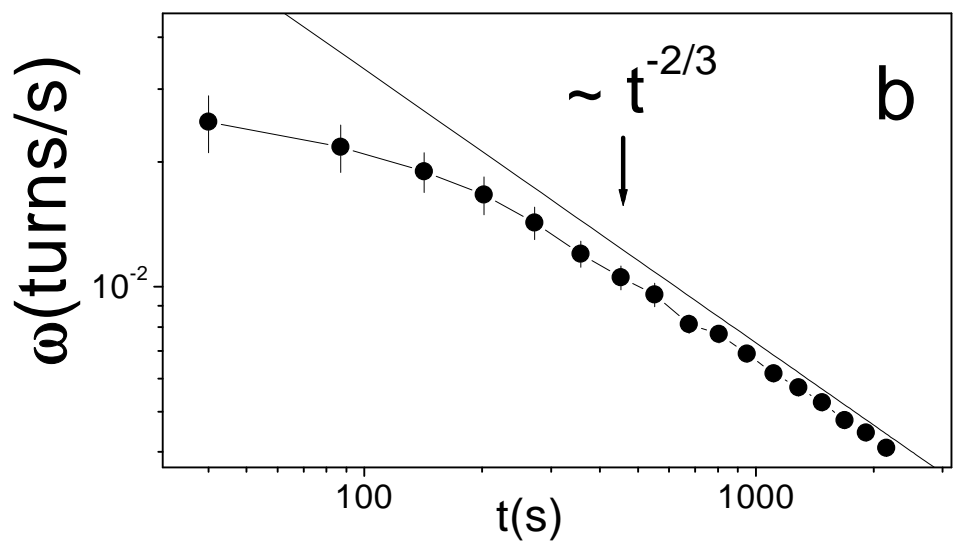
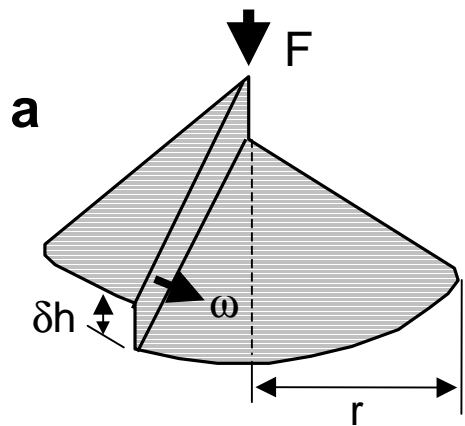
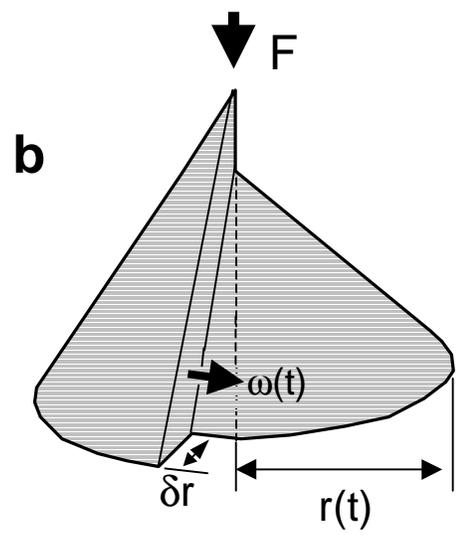
\end{document}